\documentstyle[aps,preprint]{revtex}
\begin{document}
\title{Kardar-Parisi-Zhang model for the fractal structure of cumulus cloud fields}
\author{Jon D. Pelletier}
\address{Department of Geological Sciences, Snee Hall, Cornell University,
Ithaca, NY 14853}
\maketitle

\begin{abstract}
We model the ascent of warm, moist air in the Earth's atmosphere by  
turbulent convection and expansion with the KPZ equation, familiar in 
the physics literature on surface growth. Clouds form in domains
where the interface
between the rising air and its surrounding air achieves an elevation
higher than that necessary for condensation. 
The model predictions are consistent with the perimeter 
fractal dimension and the cumulative
frequency-size distribution of cumulus 
cloud fields observed from space. 
\end{abstract}
\vspace{1cm}

\narrowtext
In a pioneering study, Lovejoy \cite{lovejoy} computed the fractal dimension of the 
perimeter of rain and cloud areas from scales of 1 to 1000 kilometers 
to be $1.35\pm 0.05$. Rys and Waldvogel \cite{rys} carried out the 
same analysis to characterize the shape of hail clouds. 
For scales above 3 km they obtained a fractal dimension consistent
with Lovejoy's result. At scales below 3 km, the authors found that severely 
convective hail storms have perimeters with the usual Euclidean dimension of 1.
Cahalan and Joseph \cite{cahalan} and Zhu et al. \cite{zhu} 
extended their methodology, including the calculation of cumulative frequency-size
distributions of cumulus cloud
fields. They found 
cumulative frequency-size distributions, the number of clouds greater
than or equal to an area $A$,
to be a power-law function of area with an exponent close to $-1$ for
some cumulus cloud scences up to spatial scales of 10 km. 

Two models have been proposed to explain aspects of the fractal structure 
of cumulus cloud fields. Hentschel and Procaccia \cite{hentschel} have considered the
turbulent mixing of an initially compact cloud using a 
theory of turbulent diffusion to explain Lovejoy's result. 
Their model does not appear
to favor any particular cloud size distribution.
Nagel and Raschke \cite{nagel} have proposed a cellular
automaton model of the atmosphere as a lattice of particles subject to 
a bouyant uplift upon the initiation of condensation and a nearest 
neighbor interaction to model entrainment of fluid by a nearby updraft.
They were able to match Lovejoy's result, but only for a particular percentage
of cloud cover.
Both papers model cloud dynamics only
after the onset of condensation. It may be essential to model the 
dynamics of the ascending warm, moist air 
(and the descending air which replaces it) prior to condensation
to explain the scaling of
cumulus cloud fields up to scales of 1000 km as observed. 
Studies solving the equations of fluid motion have been
applied to the problem of cumulus cloud formation \cite{hill} but are of
too limited a spatial bandwidth to address the observed scale-invariance.

In this paper we apply a nonlinear stochastic differential
equation known in the physics literature on surface growth as the Kardar-Parisi-Zhang
(KPZ) equation \cite{kpz} to model the interface between the ascending air and
its surrounding air. The model incorporates the expansion (contraction) of
ascending (descending) air, its random turbulent convection, and  
the entrainment of fluid by a nearby updraft. Clouds form in those
domains where the interface lies above some threshold elevation. We use the
results of Kondev and Henley \cite{kondev} on the perimeter fractal dimension
and size distribution of contour loops of random Gaussian surfaces to relate
the KPZ Hausdorff measure to the observables of cumulus cloud fields.  

We will apply the KPZ equation to the evolution of a thin fluid layer, 
originally horizontal, in the atmosphere.
Two principal processes act on a warm, moist air mass in the atmosphere:  
1) as the air is heated from below with long-wavelength outgoing radiation,
convective instabilities transport the air vertically
and 2) expansion occurs as ascending air enters regions of lower
pressure higher in the atmosphere. 
Since it is impossible to determine where the convective
instabilities will develop,
a stochastic model for this transport is appropriate.
We will model the force of convective instabilites on the fluid layer as a
Gaussian white noise. A Gaussian white noise
force, combined with the drag induced on the fluid layer by the
air it displaces, is consistent with the observed Gaussian, uncorrelated
(above the Lagrangian timescale: on the order of minutes in the atmosphere)
velocity
fluctuations in a stably or neutrally stratified atmosphere \cite{lumley}. 
The viscosity of air results in a shear force between an updraft
and nearby air which results in an effective surface tension of the
fluid layer. 
The force of convection and effective surface tension will result 
in a vertical velocity of the fluid layer (since the forces are balanced 
by the drag exerted by the adjacent fluid)  
based on these parameterizations as
\begin{equation}
\frac{\partial h}{\partial t}=\nu \nabla ^{2}h+\eta(x,y,t)
\end{equation}
where $h$ is the elevation of the layer and $\eta (x,y,t)$ is Gaussian
white noise. 

In addition to the convective transport, 
the pressure gradient with height causes ascending (descending) air
to expand (contract). 
The simplest model of this
expansion and contraction is a constant growth of the interface directed 
everywhere perpendicular to the interface with a non-zero upward component for
the layer as a whole denoted by $r$. 
This model corresponds to a constant
pressure difference between the ascending air and the air above it. 
The local vertical 
component of growth is 
equal to $r(1+(\nabla h)^{2})^{\frac{1}{2}}$. If we assume
that the gradients of the interface are small, or if we compare our model 
to only large-scale structure, we can approximate this expression as
$r+\frac{r}{2}(\nabla h)^{2}$. This Taylor expansion procedure is the same 
formulation employed by 
Kardar, Parisi, and Zhang \cite{kpz} to motivate the nonlinear term $(\nabla h)^{2}$
to model lateral growth on atomic surfaces.
The resulting differential equation for the height of the interface is
\begin{equation}
\frac{\partial h}{\partial t}=\nu\nabla ^{2}h+r+\frac{r}{2}(\nabla h)^{2}+
\eta (x,y,t)
\end{equation}
This is the KPZ
equation.

The KPZ equation with a two-dimensional surface has been solved numerically by
Amar and Family \cite{amar} and Moser, Wolf and Kertesz \cite{moser}. 
The solution is a surface with a Gaussian distribution
of elevations and a variance which depends upon the linear size of the surface
as $V\propto L^{2H}$ where $H\approx 0.4$ is known as the Hausdorff measure or
roughness exponent. 

Kondev and Henley \cite{kondev} have obtained the relationship between the fractal 
dimension of a contour loop of a Gaussian surface, $D$, and its Hausdorff measure as
$D=1.5-\frac{H}{2}$. A contour loop is a connected subset of a surface with equal elevation. 
Since clouds form above a threshold elevation where condensation begins, their base
perimeters, observable from satellite images as in Lovejoy's work,
may be 
associated with the contour loops of Kondev and Henley. Their relation,
together with the Hausdorff measure $H=0.4$, predicts
a cloud perimeter fractal dimension of $1.3$, consistent with the value 
$1.35 \pm 0.05$ observed by Lovejoy \cite{lovejoy} and Rys and Waldvogel
\cite{rys}.

In addition, Kondev and Henley have given the size distribution of contour 
lengths (the probability that a randomly chosen contour loop has a length $s$)
as $N(s)\propto s^{-\tau}$ where $\tau =1+\frac{2-H}{D}$. The cumulative 
distribution (the number of contours with length greater than $s$) 
is the integral of the noncumulative distribution,
$N(>s)\propto s^{-\frac{2-H}{D}}$. Since the length of a contour is
related to the area it encloses by $s\propto A^{\frac{D}{2}}$ (by definition), the
cumulative distribution of areas enclosed by contours is 
$N(>A)\propto A^{-\frac{2-H}{2}}$. For the KPZ Hausdorff measure of $H=0.4$
this gives $N(>A)\propto A^{-0.8}$.

In order to test the model predictions of the cumulative frequency-size distribution
against
cumulus cloud fields, we obtained global composite images from the GOES
satellites prepared at the Space Science and Engineering Center
at the University of Wisconsin, Madison for five days each in the months
of October, 1995 and January, 1996. 
The days were each separated by at least three days to ensure that each
scene was distinct. We analyzed cloud images only within 30 degrees 
latitude of the equator. Tropical clouds are ideal for study since they 
form in environments which are nearly uniform horizontally \cite{houze}. 
We divided each global scene into $60^{o}$x$60^{o}$ scenes centered
on South America, Africa, and the Western Pacific Ocean (regions of 
consistent large-scale cloud cover). To analyze smaller scales, we obtained
images of the Earth photographed from the space shuttle. We 
analyzed 16 STS-67 
images that satisfied the following criteria: 1) considerable cumulus cloud cover
untainted by other types of clouds, 
2) adequate contrast to define
cloud shapes easily,  
3) clouds were not conspicuously correlated with topography, and 4) clouds
were photographed
at a small look angle. Otherwise, the choice of the images was random. 
The resolution cell size of each image type was determined
through calibration with respect to a recognizable geographic shape. The 
resolution cell sizes of the GOES composite and shuttle images were estimated
to be 8100 km$^{2}$ and 0.084 km$^{2}$, respectively. 
We converted each image to a binary black and white image by
making all pixels darker than a certain threshold black and all those lighter than                                                                  
the threshold white. All white areas are defined as clouds for our analysis.
The cumulative 
frequency-size distribution of each image was computed and averaged
with other images of its type at equal cloud numbers. 
Least-squares linear fits to the
logarithms of the data averaged in logarithmically-spaced bins 
(so that the data was uniformly weighted in log space) yielded 
power-law exponents -0.72 and -0.82 for the GOES global composite 
and space shuttle images, respectively. These exponents are similar 
to the predicted exponent -0.8 based on the KPZ model.
In order to compare the distributions of the two types of images,
the cloud numbers for the space shuttle images were multiplied 
by a correction factor, discussed in Lovejoy \cite{lovejoy}, of the ratio of
the GOES to the space shuttle resolution cell size to the 0.8 power. 
The average cumulative frequency-size distribution for the space
shuttle images scaled in this way is plotted with the 
average
distribution of the GOES images in Figure 2 along with a least-squares
fit to the data with an exponent of -0.8. 

We have shown that a simplified model describing the dynamics of ascending warm, moist
air expanding and undergoing turbulent convection
predicts a fractal structure for
cumulus cloud fields consistent with observations of cumulus cloud
fields based upon the perimeter fractal dimension and the cumulative frequency-size 
distribution.

\begin{figure}
\caption[fig1]{}
Average cumulative frequency-size distribution, the number
of clouds greater than or equal to and area $A$, of
GOES global composite and (appropriately scaled)
space shuttle cloud images. The distributions are consistent with
the KPZ model prediction $N(>A)\propto A^{-0.81}$.
\end{figure}
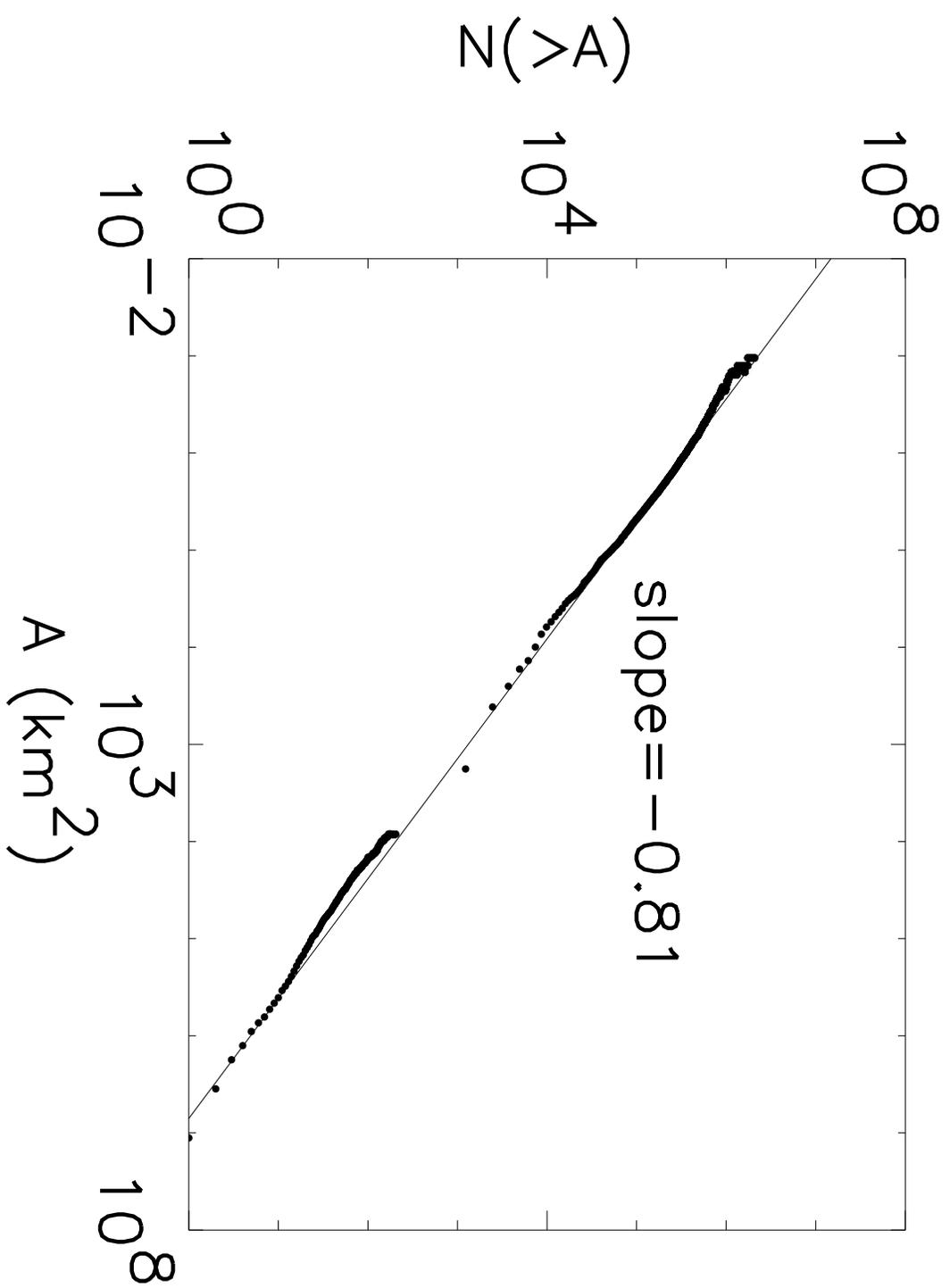
\end{document}